# Designing Laboratories for Online Instruction using the iOLab Device.

Louis Leblond[a] and Melissa Hicks[b]
[a]Associate Teaching Professor of Physics, Department of Physics, Penn State University lul29@psu.edu
[b]Director, Instructional Designer, Office of Digital Learning, Penn State University, mjs100@psu.edu

Scientific laboratories are among the most challenging course components to integrate into online instruction. Available technology restricts the design and nature of experiments and it can be hard to replicate the collaborative lab environment where frequent and immediate instructor feedback is the norm. Here we report on technological and pedagogical aspects of newly developed labs for online courses using the Interactive Online Lab (iOLab) device. We argue that this technology, coupled with an online course design emphasizing teamwork, targeted feedbacks, and self-regulation skills, provides a robust framework for students to do reliable, engaging, inquiry-based and hands-on labs outside the classroom. After describing the implementation and technology, we explain our lab objectives and how the labs were integrated into two introductory physics courses. We conclude with an example lab on kinematics.

1. **Implementation and Demographics**

Our new lab activities were created for calculus-based introductory mechanics and introductory electromagnetism (E&M) courses at the Pennsylvania State University, which are taught in an integrated lecture-lab-recitation model. The labs were implemented and revised over four years in our physics major section before being adapted to the online format. During the pilot phase, we revised the goals and objectives of the lab component of the courses [1]. Our latest implementation emphasizes observational experimental skills and critical thinking inspired by the ISLE framework of [2] together with [3].

There has been much previous work on creating labs for distance learning either with virtual simulations, hands-on equipment or by remotely operating lab equipment [4, 5, 6, 7, 8, 9, 10, 11]. In our implementation, we opted for a hands-on experience with some virtual simulations mixed in. See the work of [12, 13] for a study of the effectiveness of hands-on lab in a similar educational setting.

The online courses were first launched in Summer 2018 (mechanics) and Spring 2019 (electromagnetism) with enrollment of about 50 students per semester in each course and they have been taught every semester since then. Students are almost all software engineering majors since this is the only engineering undergraduate major offered fully online at Penn State. Students in the online courses are predominantly adult learners, many are working or have young children at home. In a recent demographics survey for the mechanics course (N = 43 students, Spring 2020 semester), we found an average age of 27 years old with one third of the students in the typical college age (18-23 years old) while the rest were older ranging from 23 to 60 years old. Students were mostly male (80%). The majority were working, with 70% reporting working full time, 15% working part time and 15% reporting that they were full time students.

2. The Technology

The iOLab device was developed by Mats Selen and Tim Stelzer at the University of Illinois at Urbana-Champaign. As of 2019, it is sold by McMillan at the cost of $199 [14] or rented (from $42 for 3 months). The iOLab is a handheld device on wheels equipped with a variety of sensors communicating wirelessly to a USB dongle. The device ships in a box that includes springs and hooks that can be attached to a force sensor. The free iOLab software collects and presents the data in real time. For electromagnetism, we supplemented the iOLab with the desktop experiment kit EM-8675 from PASCO ($50) [15]. The main iOLab features we have used are shown in Figure 1.

We have investigated multiple other options for lab equipment and found that the iOLab is the best for our needs. It is cheaper and much more versatile than third-party lab kits (e.g., Hands-on Labs, Carolina Labs, or e-Science Labs [16]). The iOLab uses real-time digital data acquisition which is an extremely desirable feature for a course at this level. The iOLab is more expensive than smartphone apps, such as Lab4U, Smart-Physics, the Physics Toolbox Sensor suite or Phyphox [17], but it has many more sensors such as the wheel, force, and voltage input sensors. Pocketlab [18] is another option similar to the iOLab without quite the same versatility.

PASCO sells a wireless smart cart equipped with the four essential sensors for mechanics. It is comparable in price to the iOLab except that the capstone software comes at an additional cost. The PASCO smart cart has much less friction than the iOLab, which can be a desirable feature for certain labs. It does not ship with a spring and, more importantly, it does not have any of the sensors needed for the subsequent course on electricity and magnetism.

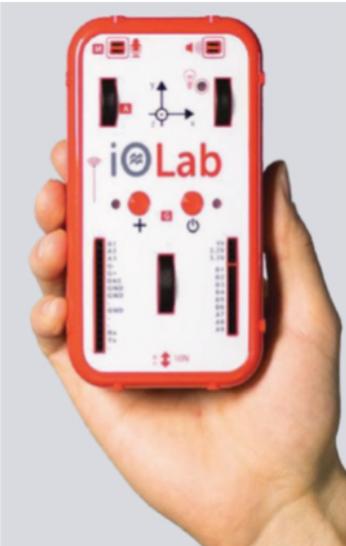

*Figure 1 iOLab sensors used in our design of the mechanics and E&M labs. We also have an exploratory lab to look at the light/microphone/thermometer and barometer sensors even though we do not use them in any other labs.*

The iOLab software is intuitive and displays data well. It has a built-in statistical tool to measure slope and area under the curve. Most of our data analysis requires scatter plots and box plots that our students create in Excel or Statkey [19] from sample data points read off the iOLab software.

Table 1 lists the labs we created together with the iOLab sensors used. The semester is 15 weeks and we usually have 8 to 11 labs after accounting for exam weeks. Four of the labs in the electromagnetism course use only equipment from the electricity kit or PhET simulations [20]. The iOLab has inspired many innovative experiments in various contexts (see example posters from the session EF, AAPT 2017 summer conference available in the Google repository found at [14]) and even more labs are possible with extra equipment. For example, Nair and Sawtell showed how the iOLab can be used to do a lab on equipotentials [21]. Bodegom, Jensen and Sokoloff have shown how to effectively adapt Real Time Physics [22] to distance learning with the iOLab [11].

| Lab | Mechanics - title | iOLab sensors used | E&M - title | iOLab sensors used |
| --- | --- | --- | --- | --- |
| 1 | Exploration | wheel | Static Electricity | - |
| 2 | Push and slide | accelerometer | Electric Field | - |
| 3 | Friction 1 | Force, accelerometer | Capacitor | analog input, battery |
| 4 | Friction 2 | Force, accelerometer | Resistance | - |
| 5 | Hooke's 1 | Force, accelerometer | RC circuit | Analog input |
| 6 | Hooke's 2 | Force, accelerometer | Magnetic Field | Magnetometer |
| 7 | Rotation | Force, accelerometer, gyro | Faraday's Law | Magnetometer high gain amplifier |
| 8 | Exploration | Barometer, sound, light | RLC circuit | - |

Table 1: The table shows the iOlab sensor(s) used in each lab activities. A few of the electromagnetism labs do not use the iOLab relying instead on the electricity kit from PASCO or PhET simulations.

Overall, the iOLab is an all-in-one tool that is powerful, portable and easy to use. For mechanics, nothing else is really needed, although students do report "iOLab fatigue" if no other equipment is used throughout the semester. They enjoy the variety that the electricity kit provides in the E&M course.

3. Lab Goals and the Course Design

Our lab objectives focus on data sense-making and observational experimental skills influenced by the work of Bonn, Holmes and Wieman [3] together with the work of Etkina and Van Heuvelen [2].

**Lab learning objectives**

By performing and discussing the lab, students will be able to:

1. present data using graphs and tables that are correctly labeled and correctly scaled.
2. design and perform a reliable experiment that investigates a phenomenon.
3. identify dependent, independent and controlled variables in an experiment.
4. identify patterns in data and devise an explanation for an observed pattern.
5. use visual tools such as box plots and scatter plots to visualize data.
6. use statistical concepts like correlation, means and outliers to make decisions based on data.
7. identify shortcomings of an experiment and suggest improvements.
8. use the slope and intercept of the "best fit line" to find the values of physical quantities.

Our objectives align with objectives 1-3 of ref [23] and the broad curricular goals of the 2014 AAPT report [1] but we emphasized depth over breadth. Our emphasis is on data sense-making using visual tools such as box and scatter plots. The web-based application StatKey [19] assisted students in the computation of all relevant statistics and provided a visual representation of the data.

The online class is structured around weekly units that include multiple assignments due throughout the week (see Figure 2). Each weekly unit includes three phases: exploration, problem solving, and revision. In the exploratory phase, students read the textbook with supplementary web-based notes, videos, and interactive questions. The main interactions occur in the problem-solving phase via a graded discussion board and a peer-review system.

| *Days* | *Tuesday-Thursday* | *Friday* | *Weekend* | *Monday* |
|---|---|---|---|---|
| | Exploratory Phase | Problem Solving Phase | | Revision |
| *Graded assignments due* | Discussion boards | First draft of tutorial and lab<br><br>Reading quiz | Peer-reviews | Final version of tutorial and lab<br><br>Quiz |

Figure 2: Example schedule for a weekly unit of the online course.

Students must submit their first post on the discussion board by the end of the exploratory phase. The discussion then continues through the rest of the week. Students are required to discuss the lab and to reply to their team members with questions, comments, compliments and/or connections. As the discussion unfolds, students must individually submit a first draft of their lab. Peer-reviewers are then assigned, and students are asked to give feedback to two other students. Feedback from teaching assistants and the instructor is provided on the discussion board and lab throughout the week. The students submit a final version of their lab at the end of the weekly unit together with a metacognitive reflection. The lab component of the course takes about 2 hours of work for the students and a similar amount of time for the TA and instructor to give feedback to students. It should be noted though, that the labs are shorter in length than what we do in residential instruction partly because the students are mostly working individually and receiving asynchronous feedback. Overall, the online course is fairly effective, with learning gains of 0.55 on the FCI (d= 1.3) using matched data [24] in Spring 2019.

It should be noted that the online courses have a larger withdrawal rate than our residential courses, which may affect our interpretation of the FCI data. It is not immediately clear whether the large withdrawal rate is artificially boosting (or lowering) the FCI gain. Anecdotally, adult learners often withdraw because of time management, not because they necessarily struggle with the material.

Measuring and quantifying the efficacy of the labs by themselves is challenging [13, 12], and we are currently working on finding ways to assess our lab objectives beyond the graded lab reports. The labs are graded very generously with most of the grade being for participation (submitting a complete lab and giving feedback to peers). About 25% of the lab grade is based on correctness using a heuristic rubric for evaluating the whole lab and specific scientific ability rubrics for each question. In addition, we have designed lab questions to assess some of the lab objectives (particularly objective 8). The average on those questions is often lower than the exam average, but not very significantly. A more precise study is needed to fully assess these labs and to compare them with similar labs in residential instruction.

A recent class survey shows that students value the lab as an important part of the learning experience, with 66 of 71 (93%) students surveyed agreeing that the iOLab-based lab was useful to their learning. Almost all students are able to operate the iOLab and the software with only a few instructional videos. We set up one-on-one Zoom meetings for technical help when needed. In our courses, we were able to avoid technical frustrations with 64 of 71 (90%) students indicating that the iOLab was easy to use. Multiple students reported enjoying the hands-on aspect of the labs, commenting that the tool made it easier for them to "see" physics concepts in action. They mentioned that it created a bridge between the concepts they were learning in class and real-world applications. Many noted the tool was fun and motivating.

Overall, we found many more positive comments about the labs than what we typically witness in residential instruction. The change in pace and the different modality of learning that the labs provide are strongly valued by our students. More information together with the latest versions of our labs are available on our website [25]. We conclude with a brief example to illustrate the typical lab experience of our students.

4. **Example from Mechanics**

"Slowing Down" is the second lab in the mechanics course and the first introduction to the iOLab accelerometer sensor and to the StatKey box plot visualization tool. The lab begins by asking students to measure the local value of "g" directly from the accelerometer. An interesting class discussion ensues on the variability of this constant (the geographic dispersion of students makes the discussion even more interesting). Students are then asked to observe the acceleration graph of a push and free sliding motion of the iOLab on its felt pads on a horizontal surface. The lab goal is to investigate differences or similarities between small and big pushes. The exact magnitude of "small vs. big" is chosen by the students. During the lab, students measure the maximum acceleration during the push and the average acceleration during the freely sliding phase (see Figure 3).

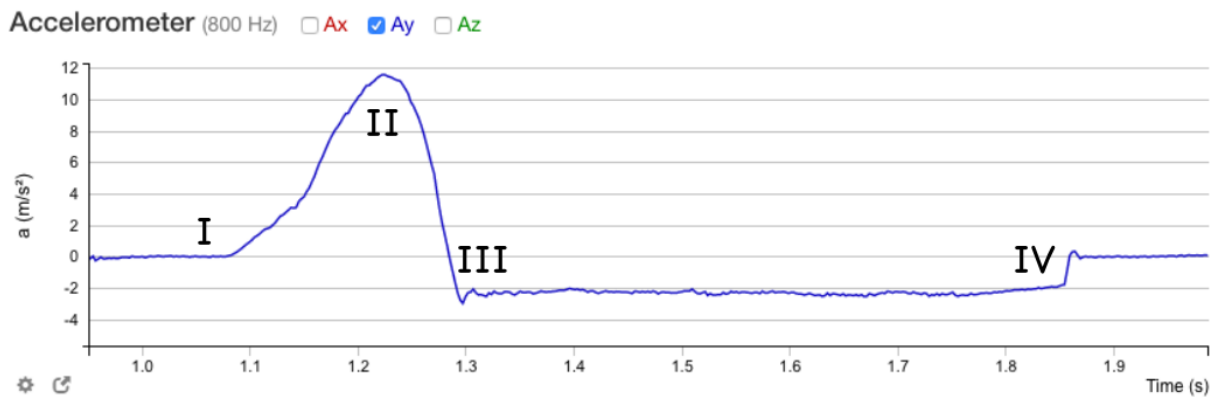

*Figure 3. Sample data from the accelerometer when pushing and letting freely slide the iOLab on a flat surface (on its felt pads). The vertical axis is the acceleration in the y direction (along the motion) and the horizontal axis is time. The push starts at point I, the iOLab reaches maximal acceleration at time II, the sliding phase starts at time III and ends at time IV. A common conceptual difficulty for students is to think that the push ends at time II even though the acceleration is still positive at that time.*

We guide the students by asking them to determine whether there is a statistically significant difference between the average acceleration during the sliding phase for a small push compared to a big push.

We do not expect our students to know or do statistical tests to determine statistical significance but, instead, we ask them to make a judgment based on looking at the box plot graph for average sliding acceleration for big versus small pushes (see Figure 4). This can be compared to a box plot for the maximal acceleration during the push where the big versus small boxes are clearly separated (not pictured).

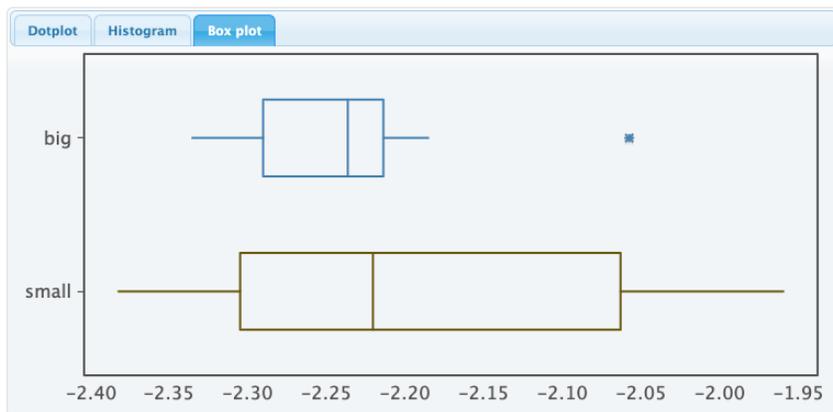

*Figure 4: Example of student box plot graph in StatKey. The vertical axis is the categorical variable big or small push. Horizontal axis is sliding acceleration in m/s². StatKey is a web application [19] that quickly provides summary statistics and a visualization of the data. Students must first create a .csv file with all relevant data and upload it to StatKey.*

The lab is open-ended. Students are not told the number of measurements they should make or even how to measure the average sliding acceleration. The lab has different questions that are either graded on correctness or graded using the Investigative Science Learning Environment (ISLE) scientific ability rubric appropriate for an observing experiment [2]. This lab assessed lab objectives 1-7 (see above) and multiple kinematics learning objectives. The lab also prepares students for the third and fourth labs on forces and kinetic friction. In this early lab, students discover experimentally that kinetic friction can be modeled as nearly independent of velocity before any discussion on forces. This example illustrates the power of a data collection device like the iOLab. Nothing is needed beside the iOLab, a computer, and a flat surface. By focusing on what the data and the graphs mean, our students learn data sense-making and observational experimental skills. Additional labs and more information are available online [25].

**Summary**

The iOlab technology, supplemented by an electricity kit, allowed us to design hands-on labs for our online version of mechanics and E&M that were both reliable and engaging. The labs are open-ended with objectives to teach and assess data sense-making and observational experimental skills. We have looked at many other technologies and we found the iOLab to be the best for designing engaging labs that assess our learning objectives.

Research shows [26] that multiple channels for discussion and targeted feedback are necessary in an online environment, and labs are no exception. The online course is designed to replicate the collaborative lab environment. Students are required to discuss the labs with their peers on discussion boards. The peer-review system allows them to critique their work and the work of others [27]. This allows for teamwork while still requiring each student to individually collect all the data, perform the entirety of the lab at home by themselves, and write their own lab reports.

Having the lab equipment at home allows students to tinker more and to learn by trial-and-error. They are forced to try harder before seeking help Accountability is even higher than in residential labs, as instructors can see each student's individual work together with their questions/help on the discussion boards.

In sum, the iOLab together with a course structure that emphasizes group work and targeted feedback provides a robust framework for delivering labs for online introductory physics. The multiple sensors with real-time data acquisition allows students to perform labs that assess data sense-making and other high-level scientific skills anywhere in the world.

**Acknowledgements**

We are thankful to Mats Selen, Tim Stelzer and Katherine Ansell[1] for many interesting discussions on the iOLab technology and pedagogy. We thank Beatrice Bonga, Jose Carpio Dumler, Jonathan Guglielmon for their help in developing and testing many of the labs. We particularly thank Eric Hudson for his help in the early implementation of our labs in the physics for majors section and for comments on this draft. We thank Eugenia Etkina, Steve Van Hook and Sarah Shandera for discussions. This work was supported by a Tombros fellowship of digital learning and by a teaching grant from the Schreyer Institute for Teaching Excellence at Penn State.

---

[1] The UIUC PER group is reforming the introductory physics labs at the University of Illinois, taken by over 2,500 students per semester, using IOLab and a similar ISLE inspired approach.